\documentclass[twocolumn,showpacs,preprintnumbers,prl,amsmath,amssymb]{revtex4}

\usepackage{graphicx}
\usepackage{dcolumn}
\usepackage{bm}

\begin{document}

\title{Soft X-ray resonant scattering study of single-crystal LaSr$_2$Mn$_2$O$_7$}

\author{H.-F. Li$^{1,4}$, Y. Su$^1$, Tapan Chatterji$^{1, 2}$, A. Nefedov$^3$, J. Persson$^1$, P. Meuffels$^1$, Y. Xiao$^1$, and Th. Br\"{u}ckel$^1$}

\affiliation{
$^1$Institut f\"{u}r Festk\"{o}rperforschung, Forschungszentrum J\"{u}lich GmbH, D-52425 J\"{u}lich, Germany\\
$^2$Institut Laue-Langevin, BP 156, F-38042 Grenoble Cedex 9, France\\
$^3$Institut f\"{u}r Experimentalphysik/Festk\"{o}rperphysik, Ruhr-Universit\"{a}t Bochum, Germany\\
$^4$Ames Laboratory and Department of Physics and Astronomy, Iowa State University, Ames, Iowa 50011, USA}

\date{\today}
\begin{abstract}

Soft X-ray resonant scattering studies at the Mn $L_{\texttt{II, III}}$- and the La $M_{\texttt{IV, V}}$- edges of single-crystal LaSr$_2$Mn$_2$O$_7$ are
reported. At low temperatures, below $T_\texttt{N} \approx 160$ K, energy scans with a fixed momentum transfer at the \emph{A}-type antiferromagnetic (0 0
1) reflection around the Mn $L_{\texttt{II, III}}$-edges with incident linear $\sigma$ and $\pi$ polarizations show strong resonant enhancements. The
splitting of the energy spectra around the Mn $L_{\texttt{II, III}}$-edges may indicate the presence of a mixed valence state, e.g., Mn$^{3+}$/Mn$^{4+}$.
The relative intensities of the resonance and the clear shoulder-feature as well as the strong incident $\sigma$ and $\pi$ polarization dependences
strongly indicate its complex electronic origin. Unexpected enhancement of the charge Bragg (0 0 2) reflection at the La $M_{\texttt{IV, V}}$-edges with
$\sigma$ polarization has been observed up to 300 K, with an anomaly appearing around the orbital-ordering transition temperature, $T_{\texttt{OO}} \approx
220$ K, suggesting a strong coupling (competition) between them.

\end{abstract}


\maketitle
\section{Introduction}

Strongly correlated electron systems such as 3\emph{d} transition-metal (TM) oxides present exciting fundamental properties, e.g., the superconducting
behavior or the extraordinary colossal magnetoresistance effect. Their rich structural, magnetic and electronic properties are governed by the interplay of
lattice, spin, charge and orbital degrees of freedom. Identification and precise characterization of the ordered phases is a prerequisite for understanding
the physical properties and unusual phenomena. Among all possible experimental techniques, resonant X-ray scattering combined with its spectroscopic
characteristics stands out due to its unique sensitivity to probe the charge/orbital ordering (CO/OO). With the application of resonant X-ray scattering in
the hard X-ray regime \cite{Ishihara2002}, significant progress on the experimental quest of the orbital degree of freedom in 3\emph{d} TM compounds has
been made. The resonant enhancement of the CO/OO super-reflections in LaSr$_2$Mn$_2$O$_7$ has been observed at the Mn \emph{K}-edge \cite{Chatterji2000,
Wilkins2003-3}. However, the resonance at the \emph{K}-edge of 3\emph{d} TMs is due to dipolar excitations from the 1\emph{s} to 4\emph{p} band and thus
only indirectly reflects the 3\emph{d} electronic states. An alternative way is the resonant soft X-ray scattering. This is a powerful technique for
directly probing the ordered phases individually. Utilizing this method, some studies \cite{Wilkins2003-4, Thomas2004, Martin2006} show strong resonant
enhancements of the expected superlattice reflections around the Mn $L_{\texttt{II, III}}$-edges $(2p \rightarrow 3d)$ which are known to be very sensitive
to the details of the 3\emph{d} electronic states, supporting direct and definitive evidences for the ordering of charge, orbital and spin degrees of
freedom in various manganites.

\begin{figure} [htl]
\centering
\includegraphics[width = 0.47\textwidth] {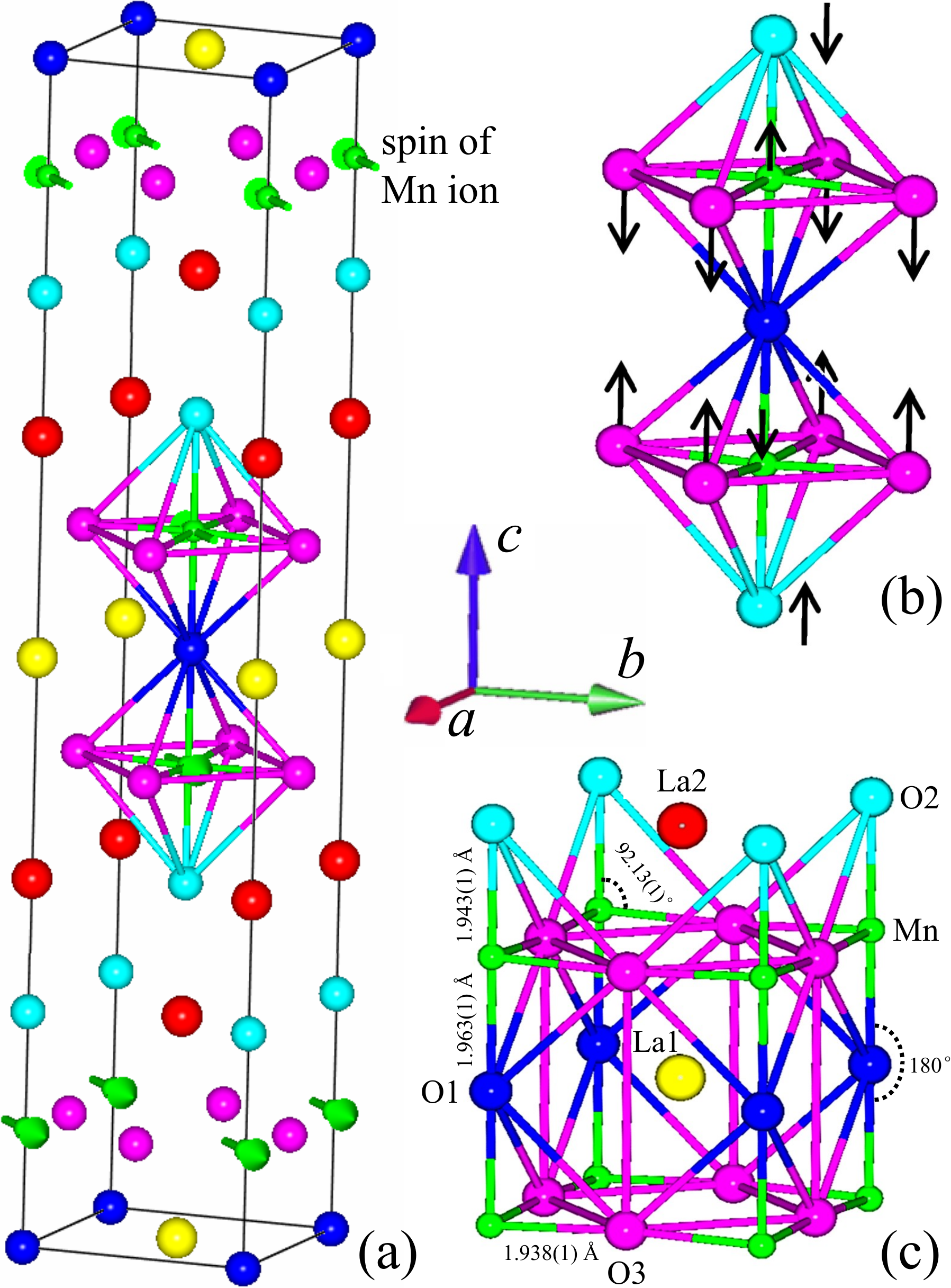}
\caption{\label{structures}(color online)
(a) Structural ($I4/mmm, Z = 2$) unit-cell of LaSr$_2$Mn$_2$O$_7$ and arrangement of \emph{A}-AFM spins of Mn ions. Structural parameters were taken from
\cite{Kubota2000}. (b) An octahedron of MnO$_6$ along the \emph{c} axis and its distortion mode. (c) Local crystal environments for the La1 and La2 sites.}
\label{structures1}
\end{figure}

The compounds La$_{2-2\texttt{x}}$Sr$_{1+2\texttt{x}}$Mn$_2$O$_7$ consist of MnO$_2$ bilayers separated by the rock-salt-type (La, Sr)$_2$O$_2$ blocking
bilayers. They are stacked vertically along the \emph{c} axis ($I4/mmm, Z=2$) as shown in Fig.\ \ref{structures}(a) taking the half-doped
LaSr$_2$Mn$_2$O$_7$ ($x = 0.5$) as an example. Though no structural transition was reported for LaSr$_2$Mn$_2$O$_7$ below room temperature
\cite{Kubota2000}, this compound undergoes a transition into the \emph{CE}-type charge-orbital ordered state at $\sim$225 K \cite{Chatterji2000}. This
state starts melting around 170 K where the \emph{A}-type antiferromagnetic (\emph{A}-AFM) structure forms [Fig.\ \ref{structures}(a)]. The (0 0 1)
reflection corresponds to this long-range \emph{A}-AFM ordering stacked along the crystallographic \emph{c} axis. The CO/OO breaks down at $\sim$100 K but
recovers again around 50 K. This reentrant behavior was discussed in terms of a polaron model \cite{Chatterji2000}. In addition, the minor \emph{CE}-AFM
phase was reported to coexist with the major \emph{A}-AFM one below $\sim$145 K and be drastically but not completely suppressed below $\sim$100 K. This
phenomenon was viewed as an effective phase separation \cite{Kubota1999}.

In this article, we report a resonant soft X-ray scattering study of the AFM (0 0 1) and the charge Bragg (0 0 2) reflections in LaSr$_2$Mn$_2$O$_7$,
varying incident photon polarizations. The detailed energy and temperature dependencies were monitored. Our study confirms that the energy spectra of the
AFM (0 0 1) over the Mn $L_{\texttt{II, III}}$-edges consist of six clear features. In addition, we found a huge resonant enhancement of the charge Bragg
(0 0 2) reflection at the La $M_{\texttt{IV, V}}$-edges.

\begin{figure} [htl]
\centering
\includegraphics[width = 0.47\textwidth] {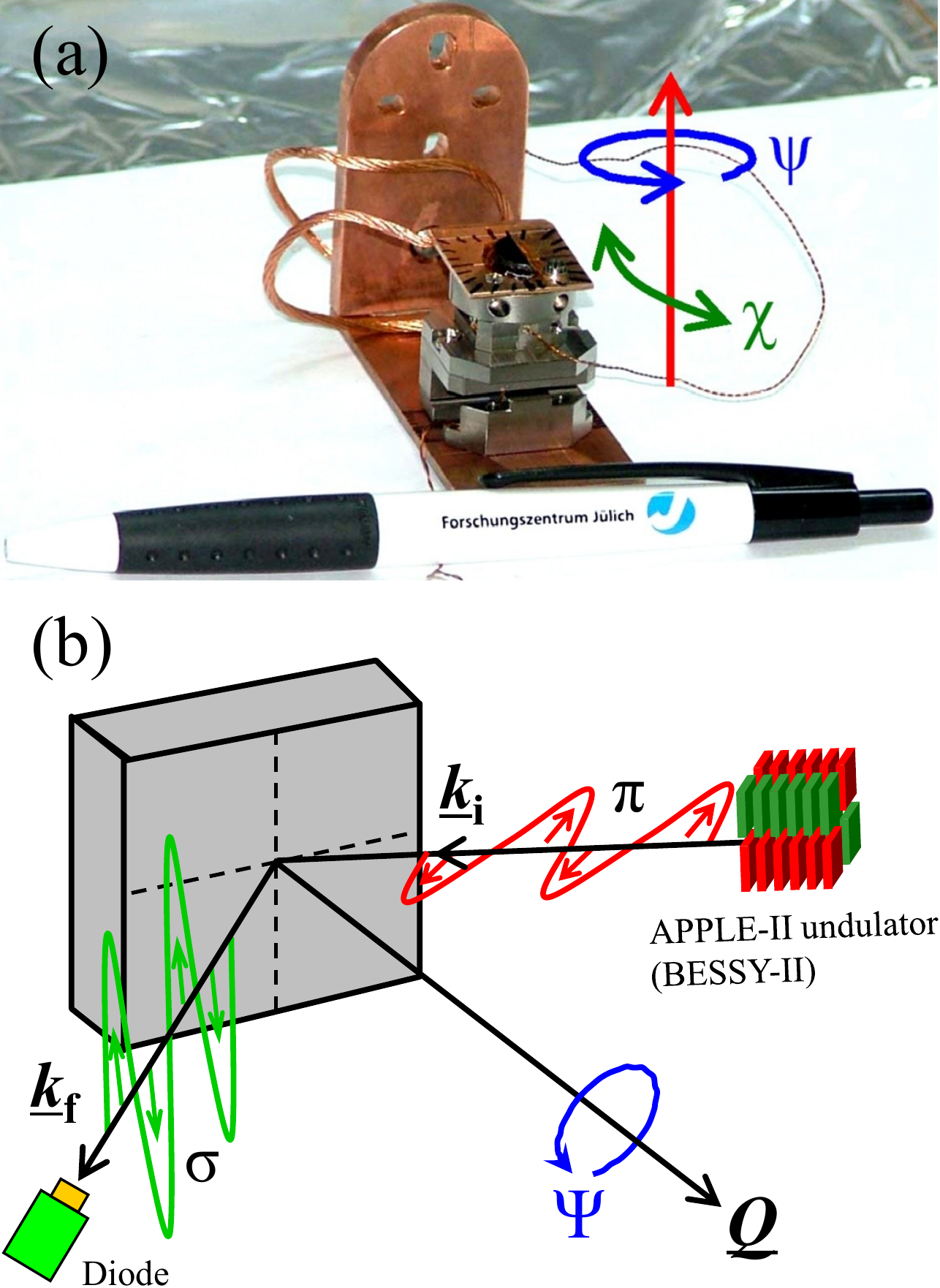}
\caption{\label{motor}(color online)
(a) Miniature goniometer utilized in this study for rotating and tilting the LaSr$_2$Mn$_2$O$_7$ sample in $\psi$ and $\chi$ angles, respectively.
(b) Basic notations for the resonant soft X-ray scattering set-up at the UHV ALICE diffractometer at Berliner Elektronen-Speicherring Gesellschaft f$\ddot{u}$r
Synchrotronstrahlung (BESSY).}
\label{motor1}
\end{figure}

\begin{figure} [htl]
\centering
\includegraphics[width = 0.47\textwidth] {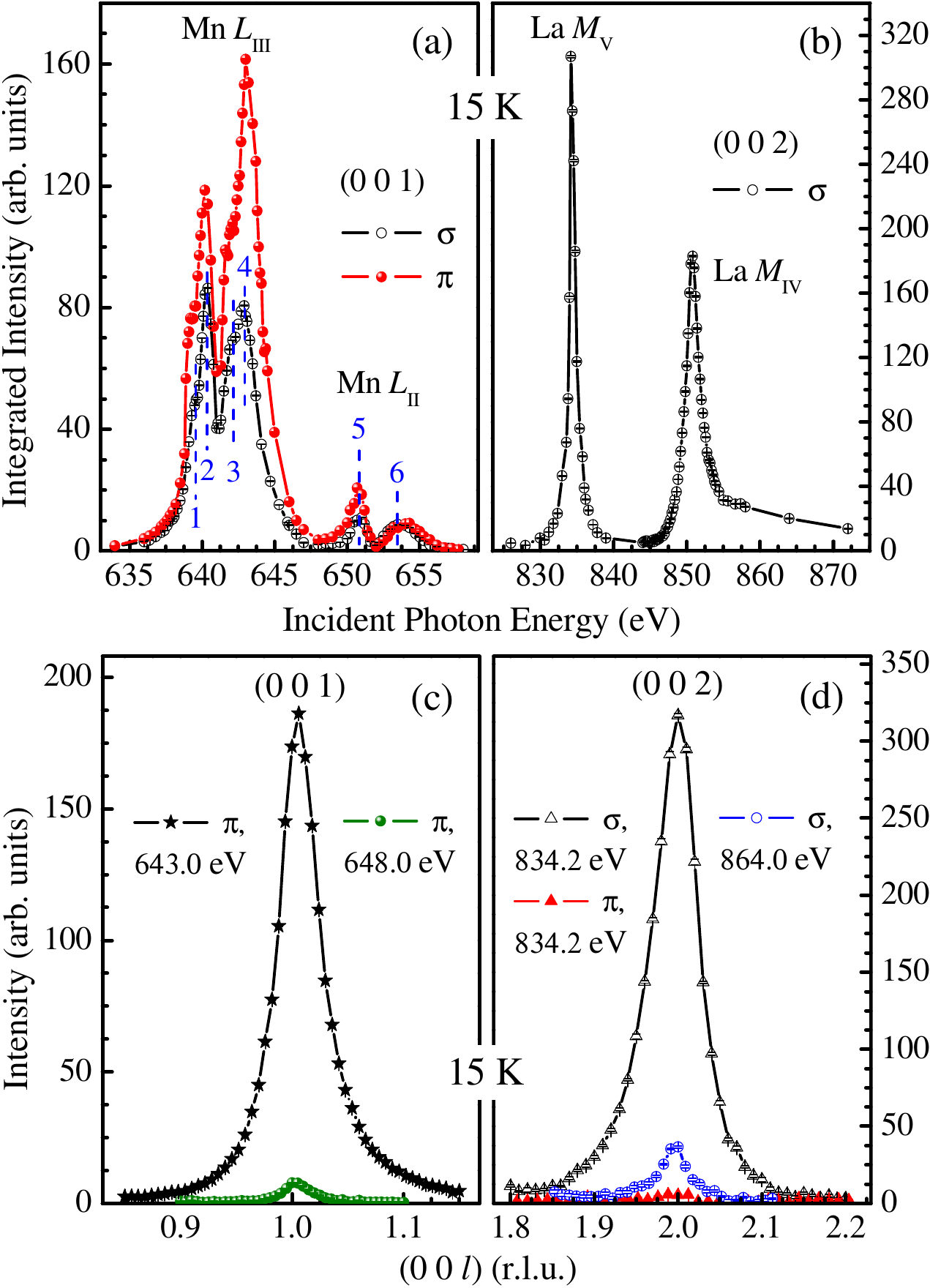}
\caption{\label{Energy15K} (color online)
Energy dependence of (a) AFM (0 0 1) recorded over the Mn $L_{\texttt{II, III}}$-edges and (b) charge Bragg (0 0 2) reflection over the
La $M_{\texttt{\texttt{IV}\texttt{, V}}}$-edges with incident linear $\sigma$ and $\pi$ polarizations at 15 K. The intensity was integrated from longitudinal
scans at each energy point. (c) and (d) longitudinal scans of the (0 0 1) and (0 0 2) reflections, respectively, at and off their strongest resonant energies
$\sim$643.0 eV (Mn $L_\texttt{III}$-edge) and $\sim$834.2 eV (La $M_\texttt{V}$-edge).}
\label{Energy15K1}
\end{figure}

\begin{figure} [htl]
\centering
\includegraphics[width = 0.47\textwidth] {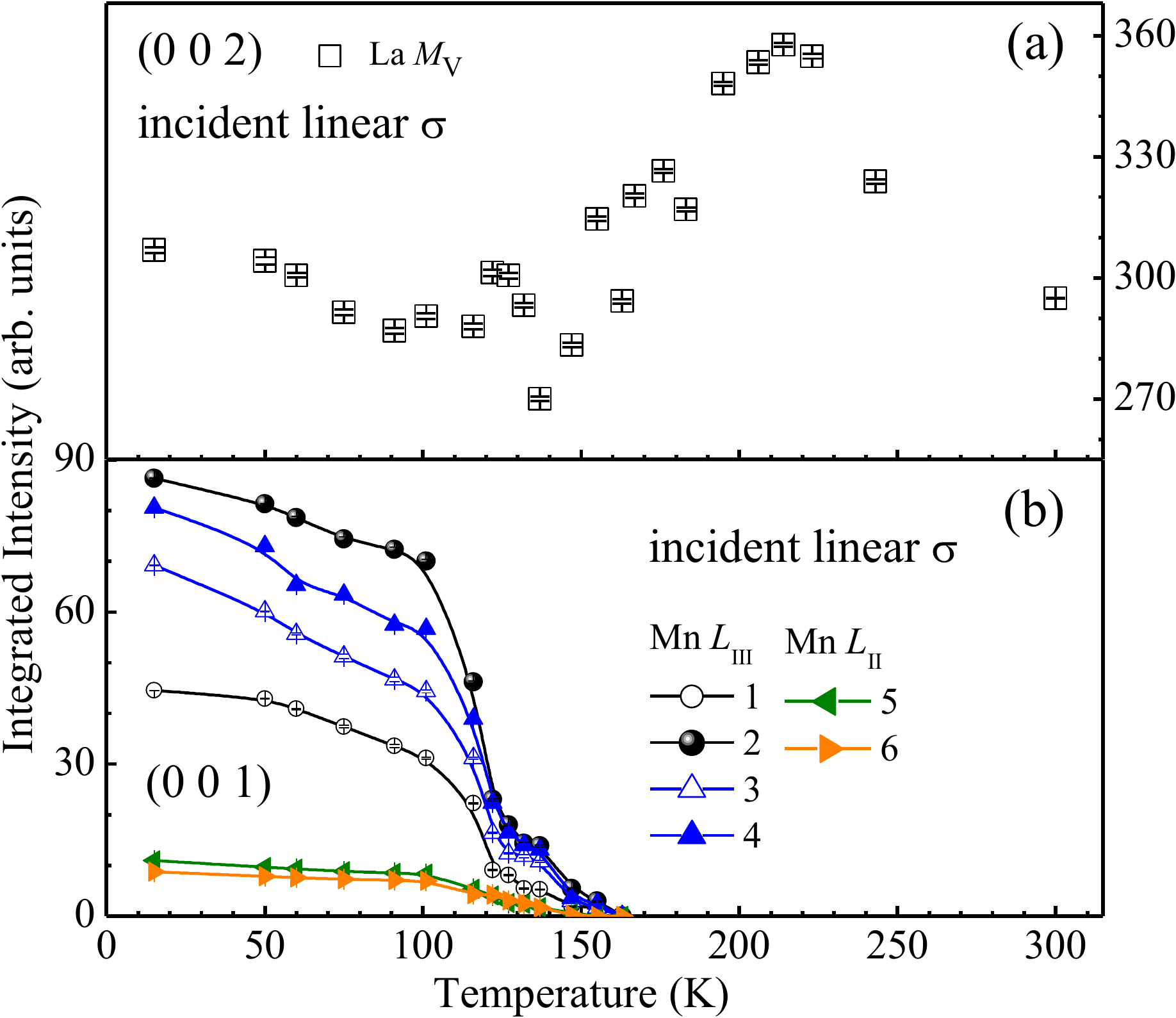}
\caption{\label{Temdepen} (color online)
Temperature dependent integrated intensity (measured in the longitudinal direction) of (a) charge Bragg (0 0 2) reflection at the La $M_{\texttt{V}}$-edges,
and (b) AFM (0 0 1) reflection recorded at the six energies corresponding to the six features 1, 2, 3, 4, 5 and 6 as labeled in Fig.\ \ref{Energy15K}(a).}
\label{Temdepen1}
\end{figure}

\begin{figure} [htl]
\centering
\includegraphics[width = 0.47\textwidth] {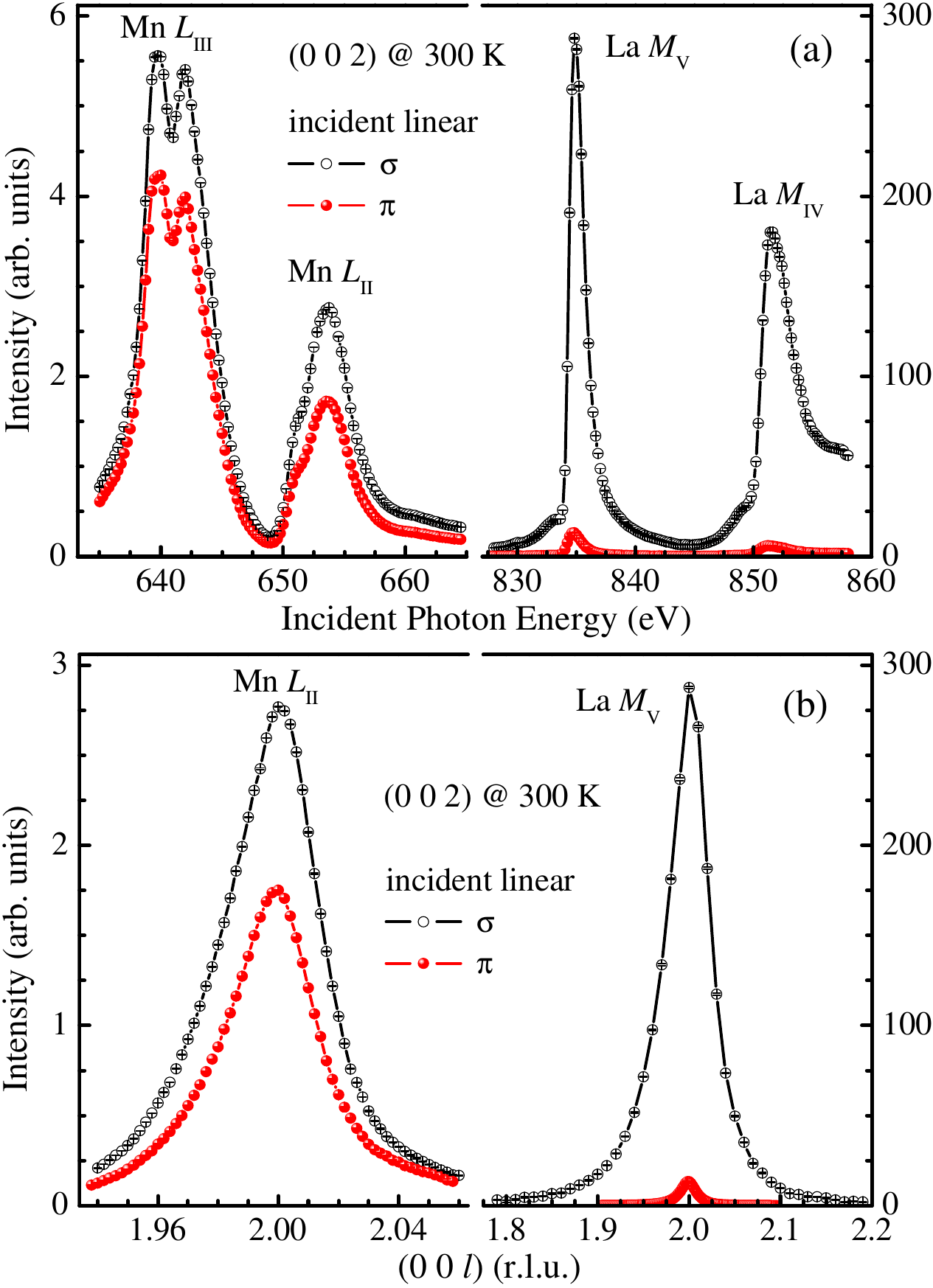}
\caption{\label{E002RT}(color online)
(a) Energy dependence of charge Bragg (0 0 2) reflection recorded over the Mn $L_{\texttt{II, III}}$- (left) and the La $M_{\texttt{IV, V}}$- (right) edges at
constant wavevector with incident linear $\sigma$ and $\pi$ polarizations at 300 K. (b) Corresponding longitudinal scans at the Mn $L_{\texttt{II}}$- and the
La $M_\texttt{V}$- edges.}
\label{E002RT1}
\end{figure}

\section{Experimental details}

Most available soft X-ray scattering chambers host two-circle diffractometers. This geometry leads to severe restrictions for sample rotating around the
surface normal ($\psi$-circle) and sample tilting ($\chi$-circle) [Fig.\ \ref{motor1}(a)], which are both essential degrees of freedom for single-crystal
resonant soft X-ray scattering studies. Therefore, a precise pre-alignment on the sample to be measured is crucial. Even when such a perfect pre-alignment
is possible, any small angular deviations that often appear during the cooling of the sample or due to the temperature-dependent change of the lattice
parameters may still easily ruin the efforts. To overcome these limitations, a new portable ultrahigh-vacuum (UHV) goniometer has been designed and built
for resonant soft X-ray scattering, which makes it feasible to adjust samples along $\chi$ ($\pm$2.5$^\texttt{o}$) and $\psi$ (360$^\texttt{o}$). This
miniature goniometer dramatically improves the efficiency of soft X-ray scattering chambers. Single crystals of LaSr$_2$Mn$_2$O$_7$ were grown by the
floating-zone method \cite{Suryanarayanan2000}. A polished one was mounted on the center of a copper plate braced to the goniometer (Fig.\ \ref{motor})
with [0 0 1] direction nearly normal to the scattering plane. Resonant soft X-ray scattering data were collected on the two-circle UHV ALICE diffractometer
\cite{Grabis2003} equipped with our goniometer at the UE56/1-PGM-b beamline of the BESSY, Germany. The vertical scattering geometry is shown in Fig.\
\ref{motor1}(b) where $\psi$ is the so-called azimuthal angle which represents the relative orientation of the sample with respect to \emph{Q}.

\section{Results and discussion}

Fig.\ \ref{Energy15K}(a) shows the energy dependence of the integrated intensity of the AFM (0 0 1) through the Mn $L_{\texttt{II, III}}$-edges at 15 K. A
similar observation on this compound was previously reported \cite{Wilkins2006} where only observed intensity without polarization analysis was recorded at
20 K. Clearly, the present study is more detailed. The spectra show a very strong polarization dependence and are dominated by scattering at the Mn
$L_\texttt{III}$-edge. Six clear features are present: four peaks (2, 4, 5 and 6) and two shoulders (1 and 3). The intensity ratio between the two main
features 2 and 4 is reversed with the different polarizations. The splitting of the energy spectra with $\sigma$ and $\pi$ polarizations around the Mn
$L_{\texttt{II, III}}$-edges is probably ascribed to the existence of a mixed valence state, e.g., Mn$^{3+}$/Mn$^{4+}$ type. The contrast intensity between
both spectra results from different orientations of the incident linear $\sigma$ and $\pi$ X-ray relative to the magnetic moments. The resonance of the
charge Bragg (0 0 2) reflection at the La $M_{\texttt{\texttt{IV}, \texttt{V}}}$-edges with $\sigma$ polarization [(Fig.\ \ref{Energy15K}(b)] was
unexpectedly observed at 15 K. This resonant enhancement is extremely large and has dramatic polarization dependence at the La $M_{\texttt{IV, V}}$-edges
as shown in Fig.\ \ref{Energy15K1}(d) where the resonance with the incident $\pi$ polarization is very small. Figures.\ \ref{Energy15K1}(c) and (d)
comparatively show the longitudinal scans of the AFM (0 0 1) and the charge Bragg (0 0 2) reflections, respectively, at and off their strongest resonant
energies $\sim$643.0 eV (Mn $L_\texttt{III}$-edge) and $\sim$834.2 eV (La $M_\texttt{V}$-edge).

The temperature dependences of the AFM (0 0 1) reflection at the six energies corresponding to the six features 1, 2, 3, 4, 5 and 6 [Fig.\
\ref{Energy15K}(a)] and the charge Bragg (0 0 2) reflection at the La $M_{\texttt{V}}$-edge with $\sigma$ polarization upon warming were shown in Fig.\
\ref{Temdepen}. The integrated intensities of the AFM (0 0 1) decrease gradually as increasing temperature to $T_\texttt{N} \sim$160 K and then disappear
simultaneously [Fig.\ \ref{Temdepen}(b)]. The intensity of the charge Bragg (0 0 2) reflection persists up to 300 K and exhibits a broad peak around 220 K
[Fig.\ \ref{Temdepen}(a)]. The recorded intensity of the charge Bragg (0 0 2) reflection over the Mn $L_{\texttt{II, III}}$- and the La $M_{\texttt{IV,
V}}$- edges at 300 K was shown in Fig.\ \ref{E002RT}(a). The corresponding longitudinal scans at the Mn $L_{\texttt{II}}$- and the La $M_\texttt{V}$- edges
were shown in Fig.\ \ref{E002RT}(b). The intensity ratio between the two main peaks at the Mn $L_{\texttt{III}}$-edge keeps the numerical relationship with
$\sigma$ and $\pi$ polarizations. Compared to Ref. \cite{Wilkins2003-4} where the La$_{1.05}$Sr$_{1.95}$Mn$_2$O$_7$ sample was investigated, the resonance
observed here has a different spectral shape and much clearer features, i.e., an obvious peak splitting at the Mn $L_{\texttt{III}}$-edge, probably
ascribing to the different Mn 3\emph{d} electronic states and the different scattering factors of the charge Bragg (0 0 2) reflection resulting from the
different doping levels. Whereas, the spectra around the La $M_{\texttt{IV, V}}$-edges involve only a single resonant peak each and this resonant
enhancement is extremely large with dramatic polarization dependence as shown in Fig.\ \ref{E002RT}. Developing a correct theoretical model to simulate the
observed energy spectra is indispensable for further understanding the physics behind.

Superlattice reflections corresponding to the propagation vector \textbf{Q} = $(\frac{1}{4} \frac{1}{4} 0)$ in LaSr$_2$Mn$_2$O$_7$ were believed to be
relevant to the 3\emph{d} OO $(3x^2-r^2/3y^2-r^2)$ of Mn$^{3+}$ ions, which is accompanied by the CO of 1:1 Mn$^{3+}$/Mn$^{4+}$ species \cite{Kimura1998}.
X-ray, neutron and electron diffraction studies \cite{Chatterji2000, Kimura1998} indicate that the coupled CO/OO developed at $\sim$225 K starts melting at
the \emph{A}-AFM transition temperature, $\sim$170 K, and collapses below $\sim$100 K. However, the CO wavevector was subsequently suggested to be
$(\frac{1}{2} \frac{1}{2} 0)$ different with that of the OO \cite{Wilkins2003-3}. The direct observation of the orbital reflection $(\frac{1}{4}
\frac{1}{4} 0)$ \cite{Wilkins2006} using resonant soft X-ray scattering technique at the Mn $L_{\texttt{II, III}}$-edges shows that the OO develops at
$\sim$225 K and persists down to $\sim$20 K, with a small change in the gradient of the observed intensity below $\sim$100 K. Comparing Refs.
\cite{Wilkins2003-3, Wilkins2006, Kimura1998}, it seems that the OO causes an occurrence of the CO and subsequently drives the formation of the long-range
\emph{A}-AFM spin ordering. The breakdown of CO is due to the ferromagnetic spin couplings in crystallographic \emph{a}-\emph{b} plane in the AFM state. In
addition, the change in the temperature-dependent integrated intensity of the charge Bragg (0 0 2) reflection at the La $M_{\texttt{V}}$-edges [Fig.\
\ref{Temdepen}(a)] seems to be inversely correlated to that of the OO reported in \cite{Wilkins2006}, indicating a possible competition between them.
Therefore, complicated cooperation and competition between spin, charge, orbital and lattice degrees of freedom exist in double-layered
LaSr$_2$Mn$_2$O$_7$.

It is interesting to explore the reasons for the huge resonant enhancement of the charge Bragg (0 0 2) reflection at the La $M_{\texttt{IV, V}}$-edges.
This peak is allowed for the tetragonal structure of this compound. The observed intensity in resonant soft X-ray scattering is mainly determined by the
anomalous atomic scattering factor of the resonant atoms. This factor is a function of the incident photon energy \emph{E} and can be expressed as $f(E) =
f_1(E) + if_2(E)$, where the imaginary part $f_2(E)$ is related to the absorption coefficient and the real part $f_1(E)$ can be deduced from $f_2(E)$
through the mutual Kramers-Kronig relation \cite{Als-Nielsen2001}. The X-ray absorption spectrum of this compound at the La $M_{\texttt{IV, V}}$-edges
needs to be measured for a quantitative calculation of the anomalous scattering factor. The anomalous atomic scattering factor shows a tensorial character
and its anisotropy is mainly related to the distortion of the local environment. During the resonant soft X-ray scattering process, in principle, the
excited electron is sensitive to any anisotropy around the absorbing ions, e.g., the anisotropy of charge, orbital, spin or lattice. There are two types of
distortion related to the MnO$_6$ octahedra in manganites: (i) one is the Jahn-Teller (JT) distortion that is inherent to the high-spin (\emph{S} = 2)
Mn$^{3+}$ ions, resulting in different Mn-O bond lengths, which is accompanied by the OO of occupied Mn 3\emph{d} orbitals; (ii) another is the cooperative
rotation corresponding to the Mn-O-Mn bond angle and tolerance factor, leading to lattice modulations and forming the octahedral tilt-ordering (TO). An
octahedral TO of Pr$_{1-x}$Ca$_x$MnO$_3$ and LaMnO$_3$ \cite{Zimmermann2001} was observed by tuning the incident X-ray energy to the $L_\texttt{I},
L_{\texttt{II}}$, and $L_{\texttt{III}}$ absorption edges of Pr and La, respectively. The structural data of LaSr$_2$Mn$_2$O$_7$ as a function of
temperature was not completed \cite{Ling2000} and the existing structural parameters were strongly challenged by the single-crystal study
\cite{Argyriou2000}. The tetragonal symmetry ($I4/mmm$) of LaSr$_2$Mn$_2$O$_7$ determines that the out-of-plane Mn-O-Mn bond angle along the \emph{c} axis
is 180$^\texttt{o}$ and the JT distortion size which is defined as the ratio of the averaged apical and the equatorial Mn-O bond lengths is close to unity
down to 10 K [Fig.\ \ref{structures}(c)]. Thus no octahedral TO exists in LaSr$_2$Mn$_2$O$_7$. However, the distortion mode developed for the MnO$_6$
octahedra based on the Mn-O bond lengths indeed exists [Fig.\ \ref{structures}(b)], which was confirmed by a Raman spectroscopy study \cite{Yamamoto2000}.
Raman spectroscopy is a very sensitive probe of the local and dynamical structural changes. This scattering method can be used to study the complex
interplay of lattice dynamics with ordering parameters in manganites. In \cite{Yamamoto2000}, the major intensity of out-of-plane spectra was assigned to
this stretching mode [Fig.\ \ref{structures}(b)] and is independent of temperature. In addition, two tiny intensities in out-of-plane spectra were
attributed to the Raman allowed modes in the original structure: atomic motion of La/Sr ions along the \emph{c} direction. Moreover, the rest appreciable
intensities were thought to be from the activated modes, i.e., the coupling between lattice distortion and CO/OO. This octahedral distortion mode has
little effect on the local crystal environment of La1 sites that locate at the center of four double-MnO$_6$ octahedra [Fig.\ \ref{structures}(c)], while
it has a profound effect on that of the La2 sites due to the asymmetric action. Indeed, based on the reported structural parameters at room temperature
\cite{Kubota2000}, the calculated local distortion sizes of La1 and La2 sites by Fullprof suite \cite{Rodriguez-Carvajal1993} are 3.2 $\times$ 10$^{-5}$
and 1.4 $\times$ 10$^{-3}$, respectively. The strongly produced anisotropy of La2 sites may lead to the resonant enhancement of the charge Bragg (0 0 2)
reflection at the La $M_{\texttt{IV, V}}$-edges. In addition, the observed competition between OO and temperature dependent integrated intensity of the
charge Bragg (0 0 2) reflection is in agreement with Ref. \cite{Yamamoto2000}.

To summarize, a systematic resonant soft X-ray scattering study at the La $M_{\texttt{IV, V}}$-edges on the possible lattice modulations and at the Mn
$L_{\texttt{II, III}}$-edges on the \emph{A}-AFM structure in LaSr$_2$Mn$_2$O$_7$ has been accomplished. At 15 K, well below $T_\texttt{N}$ $\sim$160 K,
dramatic enhancements of the charge Bragg (0 0 2) and the AFM (0 0 1) reflections at the La $M_{\texttt{IV, V}}$- and the Mn $L_{\texttt{II, III}}$- edges
were observed, respectively. The temperature dependences of the AFM (0 0 1) resonance at the six featured energies show a similar trend and disappear
simultaneously above $T_\texttt{N}$. The resonant intensity of the charge Bragg (0 0 2) reflection persists from 15 K to 300 K and was strongly strived by
the OO around 220 K.

\section{acknowledgments}

We are grateful for the excellent technical support from BESSY-II, Germany. This work was partially supported by the BMBF under contract No O3ZA6BC2.


\begin{references}

\bibitem{Ishihara2002} S. Ishihara and S. Maekawa, Rep. Prog. Phys. \textbf{65}, 561 (2002).
\bibitem{Chatterji2000} T. Chatterji, G. J. McIntyre, W. Caliebe, R. Suryanarayanan, G. Dhalenne, and A. Revcolevschi, Phys. Rev. B \textbf{61}, 570
    (2000).
\bibitem{Wilkins2003-3} S. B. Wilkins, P. D. Spencer, T. A. W. Beale, P. D. Hatton, M. v. Zimmermann, S. D. Brown, D. Prabhakaran, and A. T. Boothroyd,
    Phys. Rev. B \textbf{67}, 205110 (2003).
\bibitem{Wilkins2003-4} S. B. Wilkins, P. D. Hatton, M. D. Roper, D. Prabhakaran, and A. T. Boothroyd, Phys. Rev. Lett. \textbf{90}, 187201 (2003).
\bibitem{Thomas2004} K. J. Thomas, J. P. Hill, S. Grenier, Y-J. Kim, P. Abbamonte, L. Venema, A. Rusydi, Y. Tomioka, Y. Tokura, D. F. McMorrow, G.
    Sawatzky, and M. van Veenendaal, Phys. Rev. Lett. \textbf{92}, 237204 (2004).
\bibitem{Martin2006} J. H. Mart\'{\i}n, J. Garc\'{\i}a, G. Sub\'{\i}as, J. Blasco, M. C. S\'{a}nchez, and S. Stanescu, Phys. Rev. B \textbf{73}, 224407
    (2006).
\bibitem{Kubota2000} M. Kubota, H. Fujioka, K. Hirota, K. Ohoyama, Y. Moritomo, H. Yoshizawa, and Y. Endoh, J. Phys. Soc. Jpn. \textbf{69}, 1606 (2000).
\bibitem{Kubota1999} M. Kubota, H. Yoshizawa, Y. Moritomo, H. Fujioka, K. Hirota, and Y. Endoh, J. Phys. Soc. Jpn. \textbf{68}, 2202 (1999).
\bibitem{Suryanarayanan2000} R. Suryanarayanan, G. Dhalenne, A. Revcolevschi, W. Prellier, J. P. Renard, C. Dupas, W. Caliebe, and T. Chatterji, Solid
    State Commun. \textbf{113}, 267 (2000).
\bibitem{Grabis2003} J. Grabis, A. Nefedov, and H. Zabel, Rev. Sci. Instr. \textbf{74}, 4048 (2003).
\bibitem{Wilkins2006} S. B. Wilkins, N. Stoji\'{c}, T. A. W. Beale, N. Binggeli, P. D. Hatton, P. Bencok, S. Stanescu, J. F. Mitchell, P. Abbamonte, and M.
    Altarelli, J. Phys.: Condens. Matter \textbf{18}, L323 (2006).
\bibitem{Kimura1998} T. Kimura, R. Kumai, Y. Tokura, J. Q. Li, and Y. Matsui, Phys. Rev. B \textbf{58}, 11081 (1998).
\bibitem{Als-Nielsen2001} J. Als-Nielsen and D. McMorrow, Elements of modern X-ray physics (John Wiley and Sons, New York, 2001).
\bibitem{Zimmermann2001} M. v. Zimmermann, C. S. Nelson, Y.-J. Kim, J. P. Hill, Doon Gibbs, H. Nakao, Y. Wakabayashi, Y. Murakami, Y. Tokura, Y. Tomioka,
    T. Arima, C.-C. Kao, D. Casa, C. Venkataraman, and Th. Gog, Phys. Rev. B \textbf{64}, 064411 (2001).
\bibitem{Ling2000} C. D. Ling, J. E. Millburn, J. F. Mitchell, D. N. Argyriou, J. Linton, and H. N. Bordallo, Phys. Rev. B \textbf{62}, 15096 (2000).
\bibitem{Argyriou2000} D. N. Argyriou, H. N. Bordallo, B. J. Campbell, A. K. Cheetham, D. E. Cox, J. S. Gardner, K. Hanif, A. dos Santos, G. F. Strouse,
    Phys. Rev. B \textbf{61}, 15269 (2000).
\bibitem{Yamamoto2000} K. Yamamoto, T. Kimura, T. Ishikawa, T. Katsufuji, and Y. Tokura, Phys. Rev. B \textbf{61}, 14706 (2000).
\bibitem{Rodriguez-Carvajal1993} J. Rodr\'{\i}guez-Carvajal, Physica B \textbf{192}, 55 (1993).

\end{references}
\end{document}